# Ice-water Interface: Correlation between Structure and Dynamics

Saumyak Mukherjee[†], Sayantan Mondal[†], Biman Bagchi*

Solid State and Structural Chemistry Unit, Indian Institute of Science, Bengaluru-560012, India

*Email – bbagchi@iisc.ac.in
[†] S. Mu. and S. Mo. has contributed equally in this work

## Abstract

To comprehend the complexities of ice-water interface, we perform a study that attempts to correlate the altered dynamics of water to its perturbed structure at, and due to, the interface. The deviation from bulk values of structural and dynamical quantities at the interface are obtained by computer simulations. Water molecules are found to get exchanged between the ice-like and water-like domains of the interface with a time scale of the order of ~10 ps. To investigate the effect of interfaces in general, we study three other systems, namely (i) water between two hydrophobic nano-slabs, (ii) water at protein and (iii) DNA surfaces. In all these systems, we find that the difference from bulk properties become negligible beyond ~1 nm, with structural features converging to bulk values faster than the dynamical properties. Even in case of the latter, we find that single particle and collective properties behave differently. The approach to bulk values is rapid except for collective shell-dipole moments. We present a new insightful characterization of the surfaces by establishing a quantitative correlation between tetrahedrality order parameter ($q_{td}$) and dynamics by diffusion (D) and angular jumps (θ). In ice-water system we find that the variation of $q_{td}$, as we move from solid to liquid phase, correlated well with D. The correlation is found to be present in all the interfaces studied. Our results can be used to explain the experimental outcomes of the likes of dielectric relaxation and solvation dynamics.

## I. Introduction

Water is omnipresent in nature and often termed as the 'matrix of life'.[1-2] Decades of experimental and theoretical research have been devoted to liquid water in order to unveil its unusual structure and dynamics.[3-9] Although, by now, we have acquired an understanding of most of the behaviour of bulk water, the same is missing for interfacial water. The ubiquitous occurrences of interfacial and confined water, accompanied by oft-observed unanticipated anomalies have made it one of the central attentions in chemistry and biology.[10-17]

One of the first observations of anomalous dynamics of interfacial water originated from dielectric relaxation (DR) studies by Pethig, Grant, Mashimo and others[18-20]. They studied DR spectra of aqueous protein solutions and discovered a "mysterious" timescale of ~40 ps, in addition to two other known components ~10ps (bulk water) and ~10ns (protein rotation). The intermediate timescale was attributed to interfacial water and termed as 'delta-dispersion'. Nandi and Bagchi developed a theoretical description that considered the dynamical equilibrium between bound and free water molecules on the protein surface in order to explain the experimentally observed anomalies.[21-23] Later, a modified theory was proposed by Bhattacharyya *et al.* the exchange rate between bound water, free water and bulk water as well was included to explain the origin of the bimodal relaxation.[24] It was shown that the rate of exchange between bound and free states of water played an important role. However, the length scales of surface perturbations were not investigated.

In earlier studies, it has been shown that the water molecules that reside in the hydration layer (~3-5 Å thick) of proteins and DNA exhibit both fast and slow dynamics.[25-28] The distributions of the single particle rotational relaxation timescales can be described by a log-normal distribution.[25, 29] Because of this surface induced local heterogeneity in the dynamics of interfacial water molecules, we observe heterogeneous solvation dynamics on biomolecular surfaces.[26, 30] The timescales of solvation of natural probe tryptophan depends on its solvent accessibility and neighbouring residues. Scattering experiments reveal that the water molecules up to ~20 Å remain influenced by the presence of the protein/solute.[31]

By definition, surfaces can be broadly divided into two classes, namely hydrophobic and hydrophilic. In many cases one of course encounters mixed surfaces, for example, protein, DNA and other heterogeneous biomolecular surfaces.[32] Several chemical and biological processes are often driven by the nature of the surfaces involved. For example, hydrophobic interactions are important for protein folding and other association processes.[33-35] Early NMR experiments showed that orientations of water molecules in protein hydration layers are 3 times slower than the bulk.[36-37] However, fast water molecules can also be detected by studying single molecule orientations from MD simulations.[25, 29] DNA grooves also show similar properties.[38]

Interestingly, one observes an exponentially increasing attractive force between two hydrophobic surfaces with decreasing distance in water. This is known as the hydrophobic force law (HFL) which has been extensively studied both experimentally and theoretically.[12,

[39-47] The short range interactions are typically observed up to ~100 Å whereas a long range force might act up to ~3000 Å.[48] An exact theoretical description of such processes is quite formidable. Lum-Chandler-Weeks developed a density functional theory formalism to describe aqueous solvation of apolar solutes.[39-40] Their theory can predict the crossover and a critical length for the drying transition. Later the long range attractions have been described microscopically in terms of interactions of inwardly propagating oppositely directed polarisation of water dipoles.[44] Recently, in an MD simulation study, bimodal nature of HFL has been reported with a distance mediated crossover. It has also been shown that pressure induced structural modification has been reported inside hydrophobic aqueous confinement.[12] The effects of surface charge on the structure and dynamics of dipolar liquids and their relaxations in confinement have also been investigated.[49-50]

Apart from these interesting properties of interfacial water, an intriguing interface is generated when water stays in contact with ice.[51] The rich phase diagram of water itself being complicated,[52-53] ice-water interfaces are often associated with exotic structural and dynamic behaviours. Particularly, this system is different from the aforementioned ones because water molecules can get exchanged between the two coexistent phases. Kinetics of ice growth has been studied for this system using molecular dynamics simulations.[51] Interestingly, the rate of ice growth depends on the face of ice in contact with the liquid phase.[54-55] The properties of ice-water interfaces were studied by Karim and Haymet in the late 1980s.[56-57] However, the correlation between the structure and dynamics of interfacial water remains largely unexplored till date.

In an earlier study, one of us presented analytical expressions for the variation of self-diffusion coefficient across a liquid-crystal interface, as a function of the order parameters of liquid-solid transition.[58] The study was based on time dependent density functional theory with a free energy function derived from DFT, but with varying order parameters across the liquid-solid interface. The expressions were used to obtain the values of the order parameters in the solid phase. Good agreements were obtained for both the FCC and the BCC phases.

At this juncture, let us understand the nature of interfacial water briefly from a physical chemistry viewpoint. When water molecules interact with a surface (either hydrophilic or hydrophobic) or become confined, it loses entropy. On the other hand, it may gain considerable amount of enthalpy that arises because of the liquid-surface interactions. Additionally, water molecules thrive to preserve its extensive hydrogen bond network.

Hence, the interplay among several complex processes determines the structure and, in turn, the dynamics of interfacial water. As a result, several unique characteristics arise, for example, an enhanced self-dissociation, a decreased dielectric constant, an increased specific heat, emergence of both faster and slower than bulk dynamics etc., to name a few. It is noteworthy that the nature of air-water interface (acidic/basic) has still remained controversial. Hence, the presence/creation of an interface tremendously alters the properties of water.

Here we take the view that one can think of the surface effect on water as a perturbation. When a system is perturbed, the perturbation can propagate up to a certain length after which the bulk behaviour is observed. We term this length as a correlation length. Correlations can be either static or dynamic. However, a quantitative understanding and estimation of such correlation length scales are yet to be achieved. In our present work, we ask the following question- *How far, from the surface, do the correlations propagate*? We aim to understand the length scales that are involved with different static and dynamical quantities in liquid water.

In this article we study by computer simulations four different systems, namely (i) ice-water interface, (ii) water between two hydrophobic sheets, (iii) water on protein and (iv) water at DNA surfaces. Interestingly, we find that the perturbation from bulk properties become negligible beyond ~1 nm, with structural features converging to bulk values faster than the dynamical properties. As in our earlier work, we find that single particle and collective properties behave differently. The approach to bulk values is rapid for the single particle but less so for the collective shell-dipole moments. We present detailed quantification of structure by tetrahedrality order parameter (q) and dynamics by diffusion (D) and angular jumps ($\theta$). Importantly, we find that in the ice-water system, the variation of q (as we move from solid to liquid phase) correlated well with D. Such correlation is found to be present in all the interfaces studied. Our results can be used to explain the experimental outcomes of the likes of dielectric relaxation and solvation dynamics.

## II. System and Simulation Details
### A. Ice-water interface

In order to create an ice-water interface, we first obtain the equilibrated structures of ice and water separately. The structure of hydrogen disordered ice-$1_h$ is generated using the GenIce package developed by Matsumoto *et al.*[59] The ice is placed such that the crystallographic c-axis coincided with the Z-axis of the box. As a result, the secondary prismatic (11$\bar{2}$0) plane of ice-$1_h$ was perpendicular to the Z-axis. The box of ice so obtained has the dimension of $63 \times 59 \times 72$ Å$^3$. Next, we obtain a box of water so that the XY dimension of the box matches with the ice-box. Each box contains about ~8500 water molecules. Finally, we club the two boxes together along the z-axis to get the desired interface (**Figure 1a**). We use the TIP4P/ice model of water as it is known to efficiently reproduce the thermodynamics properties of the ice phase.[60] The system is first energy minimized. Thereafter we equilibrate the system for 5 ns in NPT conditions at T = 250 K and P = 1849 bar. The temperature and pressure of the system are chosen from the coexistence line of liquid and solid phases of water as found from the phase diagram of TIP4P/ice water model.[60] Subsequently, we carry out the production molecular dynamics run for 1 ns with 4 fs data dumping. We confirm the stability of the interface by comparing the density profiles of the system at the beginning and the end of the simulation.

### B. Rectangular nano-confinement

We have performed atomistic molecular dynamics (MD) simulations of SPC/E water[61] confined between two graphene sheets of dimensions (50×50) Å$^2$. We have kept the two hydrophobic sheets fixed at 50 Å apart along the Z-axis (**Figure 1b**). The LJ parameters, $\sigma$ and $\varepsilon$ for the carbon atoms have been chosen to be 0.34 nm and 0.09 kCal/mol respectively.[17, 62] In between the two sheets we have inserted 3,772 water molecules by maintaining the energy and overlap criteria. The system is first energy minimised and equilibrated for 5 ns in an NVT ensemble (T = 298 K). After that, we have further simulated the system for 10 ns with the same simulation parameters and dumped the trajectory every 4 fs.

### C. Aqueous protein system

For this work we have used the globular protein myoglobin. This system has been studied extensively in our earlier studies.[25, 63] We obtain the atomic coordinates of the protein from the crystal structure reported in the protein data bank (PDB id: 1MDN).[64] We use the GROMOS96 53 a6 force-filed[65] parameters for protein and SPC/E parameters for water.[61] The protein is first solvated using 56908 water molecules in a cubic box of length 120 Å. The system thus prepared is energy minimized and equilibrated for 5 ns at constant temperature (300 K) and pressure (1 bar) keeping the protein constrained. This is followed by a 10 ns equilibration with a mobile protein at a constant temperature of 300 K. We run the final production MD for 5 ns with a data dumping frequency of 100 fs. The water around the protein is divided into 1 Å bins for further analyses (**Figure 1c**).

### D. DNA water system

In this case we have first solvated a 38 base pair long B-DNA with the following sequence d(GCCGCGAGGTGTCAGGGATTGCAGCCAGCATCTCGTCG)$_2$ in a box of (18 × 6 × 6) nm$^3$ (with periodic boundaries) filled with 20,424 water molecules and 74 Na$^+$ ions to maintain the net monopole neutrality (**Figure 1d**). The initial configuration of the duplex DNA is obtained using the Nucleic Acid Builder (NAB) routine implemented in the AMBER package.[66] We use the AMBER99sb-ildn force field[67] for the DNA and TIP3P water model.[68] Earlier simulation studies have reported stable MD trajectories by using this combination of potentials.[38] After attaining minimum energy configuration, we have equilibrated the water and ions for 5 ns at constant temperature (300 K) and pressure (1 bar) (NPT) by applying position restrain on the DNA atoms followed by equilibration for another 5 ns without position restrain. We have performed final production runs in NVT ensemble (300 K) for 10 ns.

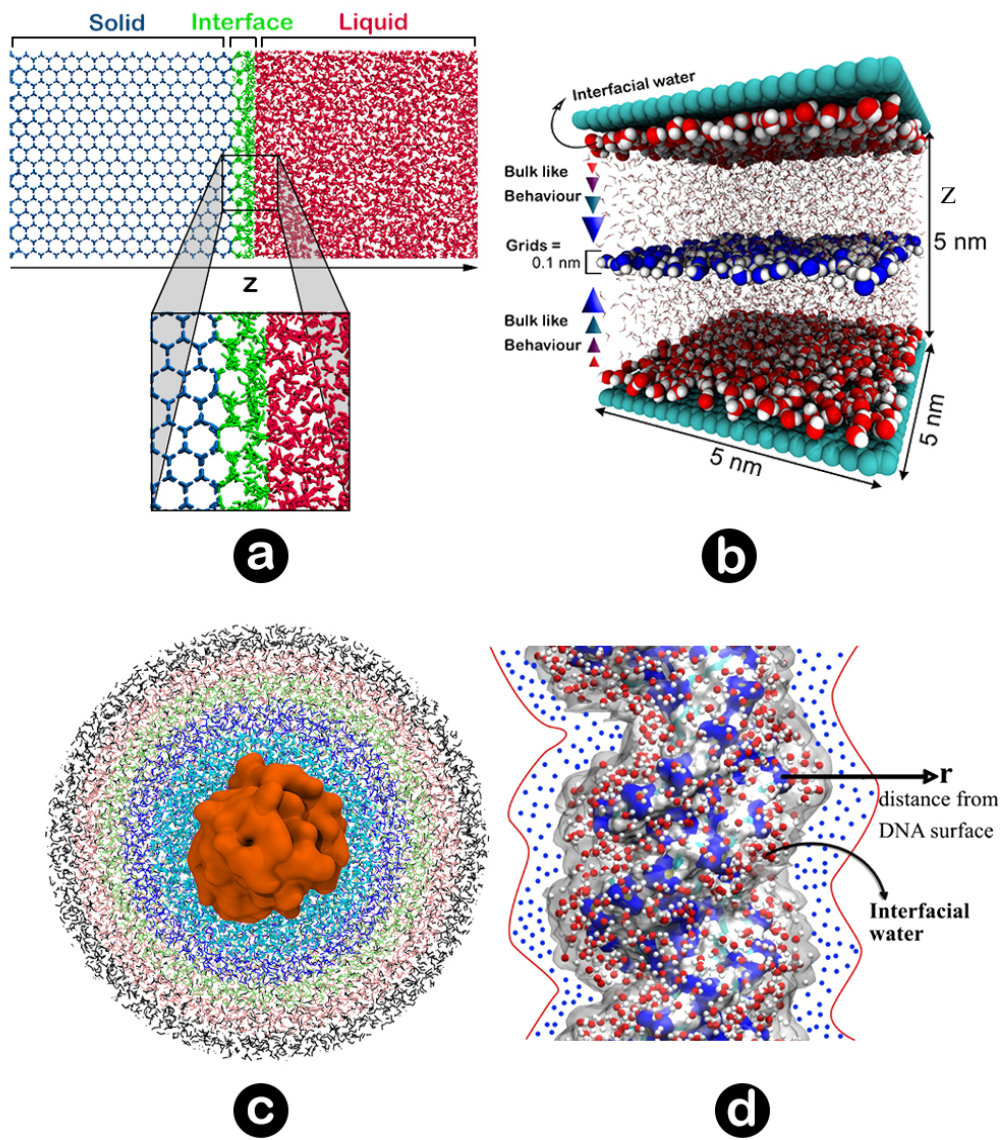

**Figure 1.** Schematic representation of the systems studied in this work. (a) Ice-water interface. The interface is zoomed out to show the arrangement of water molecules properly. A gradual transition from an ordered to a disordered state is seen. (b) Rectangular slab-confinement. The interfacial water molecules are shown in red and the bulk like water molecules are shown in blue (c) Water around the protein (Myoglobin) is divided into 1 Å spherical bins that spread out from the surface of the protein. The different layers are represented by the different colours. (d) Water at DNA surface. The water molecules are categorised in different bins of 1Å width based on their distance (r) from the DNA surface. The interfacial water molecules are shown in red and the water molecules in the next bin are shown as blue dots.

All the simulations are performed using the GROMACS v5.1.2 molecular dynamics simulation package.[69] The systems are energy minimized using a succession of steepest decent and conjugate gradient algorithms.[70] We have used leap-frog integrator[71-72] to propagate the system with dt = 2 fs. The temperature and pressure are maintained using the Nose-Hoover thermostat[73-74] with $\tau_T = 0.21\,ps$ and Parrinello-Rahman barostat[75] with $\tau_P = 2.0$ ps. We have chosen the cut-off radius for neighbour searching and non-bonded interactions to be 12 Å. We have employed LINCS algorithm[76] to constrain the bonds in the hydrophobic wall, protein and DNA. For the calculation of electrostatic interactions, we have used Particle Mesh Ewald (PME)[77] with FFT grid spacing of 1.6 Å.

## III. Theoretical Methodology

To study the length scales as discussed in the introduction, we divide each system into several grids of different shapes depending on the nature of the system and obtain the spatial variation of structural and dynamical variables. Below we detail the parameters calculated for each system.

**a) Quantification of the structure**

Water molecules, by virtue of their hydrogen bond pattern, have an inherent tendency to achieve a tetrahedral environment around them. For example, in ice-phase (ice-1h) the water molecules appear to be arranged in a completely tetrahedral geometry[78].

In the presence of hydrophobic or hydrophilic interfaces, hydrogen bond reorientations occur and the local structure gets disturbed. Hence, local tetrahedrality provides us with a definitive description of the structure in water. For a given molecule the local tetrahedrality order parameter ($q_{td}$) is defined by **Eq. [1]**.[9, 79]

$$q_{td} = 1 - \frac{3}{8} \sum_{j=1}^{3} \sum_{k=j+1}^{4} \left( \cos\psi_{jk} + \frac{1}{3} \right)^2 \quad [1]$$

Here, $\psi_{jk}$ represents the angle formed by the lines joining the oxygen atoms of the concerned water molecule and its nearest neighbour j and k ($\leq 4$). For a unit of 5 water molecules, we obtain 6 such angles. In each grid, the local tetrahedral ordering is obtained by averaging $q_{td}$ over all the molecules present in the grid. For an ideal tetrahedral arrangement ($\psi_{jk} = 109.5°$), $q_{td} = 1$. Hence, the value of this order parameter gives us an estimation of mutual arrangements among water molecules, thus revealing perturbations from interfaces.

**b) Quantification of dynamics**

We investigate two aspects of the single particle dynamical behaviour of water molecules in this work, namely translation and rotation. Translational dynamics can be quantified from the average mean square displacement (MSD, $\langle \delta r^2 \rangle$) of water molecules in a grid. It is linearly dependent on time. The slope of the curve gives the diffusion coefficient (D).

$$\langle \delta r^2 \rangle_i = 6D_i t \qquad [2]$$

Here i is the grid index. However, at ambient temperatures the dynamics of water molecules is fast. As a result, the residence times of water molecules in the grids are too low to calculate MSD that can give a reliable estimate of the diffusion coefficient. Hence for the systems other than the one containing ice, we evaluate the grid-wise translational dynamics by calculating the displacement of the water molecules within a given time window (Δt). Rotation is also evaluated by considering angular displacements within Δt time interval. This is computed by taking the arccosine of the dot product between an O-H bond vector (**b**(t)) of water at time t and (t+Δt).

$$\langle \Delta \rangle_i = \frac{1}{N_i} \sum_{j=1}^{N_i} \left( r_j(t+\Delta t) - r_j(t) \right) \qquad [3]$$

$$\langle \theta \rangle_i = \frac{1}{N_i} \sum_{j=1}^{N_i} \cos^{-1}\left( \mathbf{b}(t).\mathbf{b}(t+\Delta t) \right) \qquad [4]$$

The angular brackets denote average over the number of water molecules in a given grid i ($N_i$). However, this time gap Δt needs to be chosen judiciously. The dynamical difference in different parts of the system cannot be properly judged if the time window is too small (of the order of ~1 or ~10 fs). On the other hand, too big a window (~10 ps or more) can mask out the essential dynamical features. Hence, we calculate the grid-averaged translation and rotation suffered by the water molecules in 100 fs.

Collective dynamics is evaluated by computing the total dipole moments of the water molecules inside a grid as shown in Eq. [5].

$$\mathbf{M}_i = \sum_{j=1}^{N_i} \boldsymbol{\mu}_j \qquad [5]$$

Here, $\boldsymbol{\mu}_j$ is the dipole moment vector of $j^{th}$ water molecule.

To quantify the relaxation length scales of dynamics and structure, we evaluate the spatial correlation functions for each of the above parameters.

$$C_A(r) = \frac{\langle A(0)A(r)\rangle}{\langle A(0)^2\rangle} \quad [6]$$

For any function A(r) the spatial correlation is given by Eq. [6]. We obtain the relaxation length scales by fitting the distance dependent variations of the structural and dynamic parameters and their spatial correlations to any one of the following functions.

$$C(r) = 1 - ae^{-r/\xi} \quad [7]$$

$$C(r) = \left(1 + \frac{r}{\xi_0}\right)^{\alpha} \left\{ \sum_{i=1}^{n} a_i e^{-r/\xi_i} + \left(1 - \sum_{i=1}^{n} a_i\right) \right\} \quad [8]$$

$$C(r) = \sum_{i=1}^{n} a_i e^{-r/\xi_i} + \left(1 - \sum_{i=1}^{n} a_i\right) \quad [9]$$

$$C(r) = \left(\sum_{j=1}^{m} a_j e^{-(r/\xi_i)^2}\right) + \left(\sum_{i=1}^{n} a_i e^{-r/\xi_i}\right) + \left(1 - \sum_{i=1}^{n} a_i - \sum_{j=1}^{m} a_j\right) \quad [10]$$

$$C(r) = \left(\sum_{i=1}^{n} a_i \cos\left(\frac{r}{\xi_i^0}\right) e^{-(r/\xi_i)^2}\right) + \left(1 - \sum_{i=1}^{n} a_i\right) \quad [11]$$

$$C(r) = \left(\sum_{i=1}^{n} a_i \cos\left(\frac{r}{\xi_i^0}\right) e^{-r/\xi_i}\right) + \left(1 - \sum_{i=1}^{n} a_i\right) \quad [12]$$

$$C(r) = \left(\sum_{j=1}^{m} a_j \cos\left(\frac{r}{\xi_i^0}\right) e^{-(r/\xi_i)^2}\right) + \left(\sum_{i=1}^{n} a_i \cos\left(\frac{r}{\xi_i^0}\right) e^{-r/\xi_i}\right) + \left(1 - \sum_{i=1}^{n} a_i - \sum_{j=1}^{m} a_j\right) \quad [13]$$

It is to be noted that the last parts of Eq. [8] to [13] are off-sets that have been used for proper fitting of the data.

## IV. Results and Discussion

### A. Ice-water interface

Ice-water interface provides us with a unique system to study interfacial anomalies in water. Melting and freezing of ice has been studied extensively because of the intriguing interplay of structure and dynamics during the processes.[80-81] **Figure 2** shows the density profile of this system along with an illustrative snapshot of the coexistent solid and liquid phases in the background. The system in this representation is viewed from the basal plane (0001) of ice and the secondary prismatic (11$\bar{2}$0) plane is placed at the solid-liquid interface. We consider the direction perpendicular to the interfacial plane as the z-axis.

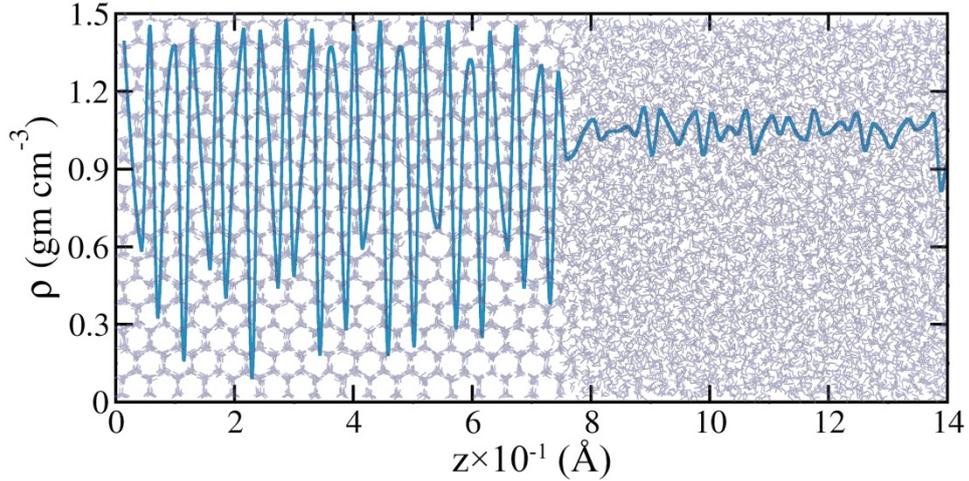

**Figure 2.** Density profile of a system of ice and water in coexistence from the basal plane of ice-$1_h$. The said system is shown as a watermark in the background. The regular arrangement of water molecules in ice phase gives rise to the large oscillations in density profile depending on the number of molecules in the grids along the z-direction. Beyond the interface, the value becomes equal to bulk water (~1 g cm$^{-3}$).

The sudden decrease in the oscillations of density denotes the interface between the two phases. The interfacial water molecules are structurally and dynamically different from the two distinctly defined phases. To characterize the structural features of the water molecules we calculate the tetrahedrality order parameter ($q_{td}$) as introduced in the previous section (**Eq. [1]**). The system is coarse grained into 14 grids along the z-axis (each 1 nm thick) and $q_{td}$ is averaged for all the water molecules in each grid. To obtain a continuous profile, we have fitted the data with a spline function.

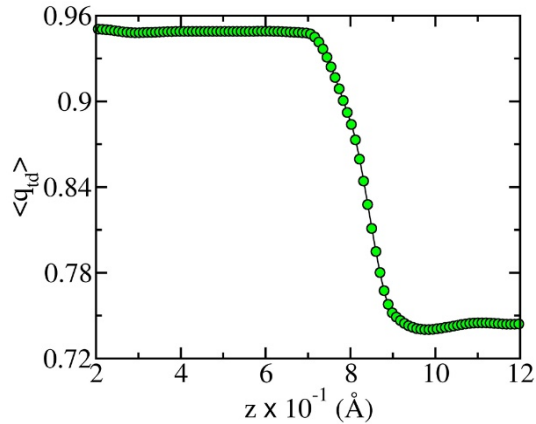

**Figure 3.** Tetrahedrality order parameter ($q_{td}$) represented as a function of z. The value of $q_{td}$ drops from ~0.95 to ~0.75 at the ice-water interface denoting severely varying structural organization on the two sides of the interface. The inflection region of the graph denotes the width of the interface (~1 nm).

For TIP4P/ice water model, the average value of $q_{td}$ is 0.95 in ice and 0.75 in water. In **Figure 3** we see that in the ice phase the value remains almost constant as we move along the positive z direction. At the interface it starts a rapid decent towards liquid bulk value. The decrease at the interface spans a distance of ~1 nm. The data is spline-fitted to get a continuous profile. The loss of structure on going from ice to water denotes a significant increase in entropy. The interfacial molecules retain the behaviour of both the phases. But how does this change in structure affect the dynamics of the system?

We find that certain water molecules get exchanged between the solid and the liquid phases. To quantify the dynamical nature of the system, we calculate the diffusion coefficient (D) of the molecules averaged in each grid according to **Eq. [2]**. Representative mean square displacements (MSD) of the different regions of the system are shown in **Figure 4a**. The gradient is almost zero in solid ice phase, denoting negligible diffusion. In liquid water, the diffusion is two orders of magnitude higher. Interfacial diffusion lies in between these two regimes.

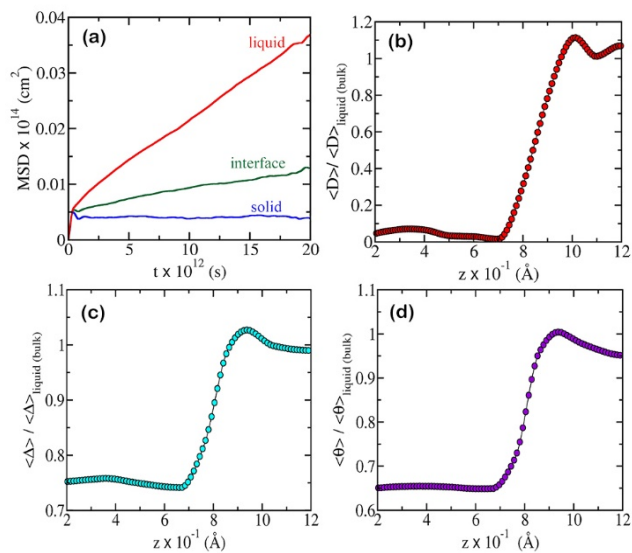

**Figure 4.** (a) Representative time evolutions of the mean square displacements of the three regions in the system, namely solid (blue), liquid (red) and the interface (green). Slope of the linear region is almost 0 for solid, while it is highest in the liquid phase. Interfacial dynamics lies in between the two extremes. (b) Diffusion coefficient D, (c) translational jump in 100 fs $\Delta$, and (d) rotational jump in 100 fs $\theta$ (normalized with respect to bulk liquid) are plotted as functions of distance along the z-direction. Abrupt increase is observed for all the variables at the interfacial region after which they attain bulk value.

To study the diffusion of water in the different parts of the system, we plot grid-averaged diffusion coefficients along the z-direction. This is shown in **Figure 4b**. The value of D is normalized by the bulk liquid value. We find that diffusion increases by more than 90 % as one proceeds from the solid to the liquid phase.

Measure of the dynamics of the system can also be quantified by determining the translational ($\Delta$) and angular ($\theta$) jumps that the water molecules execute in a given time window (**Eqs. [3]** and **[4]**). These are shown in **Figure 4c** and **d**. We see that $\Delta$ increases by ~25 % while $\theta$ increases by ~35 % on going from solid to the liquid phase. Molecular vibrations in ice allow it to retain some degree of entropy that leads to the dynamical changes. However, diffusion being a more time averaged property is almost zero in this phase. It should be noted that the distribution of structural and dynamical parameters is wider in liquid water as compared to ice. As a result, the profiles are relatively straighter in the ice phase.

An interesting aspect of the ice-water interface is its width. We find that the width depends on the variable that we use to define it, for example $q_{td}$, D etc. But how do we determine the precise location of the interface and its width? For this, we differentiate the concerned variable with respect to distance (z in this case). The inflection region of the variable leads to an extremum in the gradient giving the exact position of the interface. This formalism is known as the Gibbs dividing surface which is an imaginary line that defines the interface between two coexisting phases.[82] This is plotted for the two said variables in **Figure 5**.

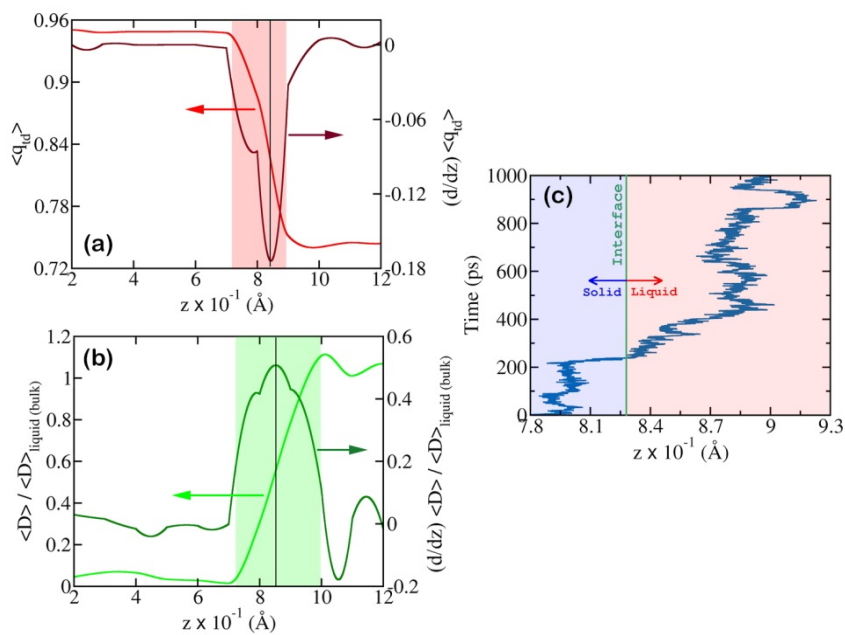

**Figure 5. Position and width of the ice-water interface is determined by plotting the gradient of (a) tetrahedrality order parameter ($q_{td}$) and (b) diffusion coefficient (D) along the z-direction. The position is denoted by the black dotted line and the width of the interface is denoted by the region within the two green lines. The dynamic interface (b) is wider than the static counterpart (a) by almost 20 Å. z-dependence of $q_{td}$ and D are also shown for reference. (c) Z-trajectory of a selected water molecule that**

**gets exchanged between ice-like to liquid-like environments across the interface. This shows the dynamic nature of the interface.**

While the position of the gradient peak denotes the mean position of the interface, its stretch is given by the width of the peak. It can be clearly seen that the interface is narrower when defined with respect to $q_{td}$ (~15 Å) than that of D (~35 Å). However, the position of the dividing surface remains almost the same. Hence, the structural parameters of the interface have a much shorter length-scale as compared to dynamical variables.

It is interesting to note that the ice-water interface is a dynamic entity. This allows water molecules to get exchanged between the liquid and the solid phases. An example of such a water molecule is shown in **Figure 5c**. It shows the z-trajectory of a selected water molecule that was initially located in the ice-like region. Time is plotted along the vertical axis. The molecule crosses the interface to enter the liquid-like environment. The timescale of this exchange process is ~10 ps.

Varying structural and dynamic length-scales in several aqueous interfaces impart intriguing characteristics to the interfacial water. Dynamics is often the consequence of structural organizations. We quantify this fact by plotting the dynamical variables of our system against the structural parameter $q_{td}$.

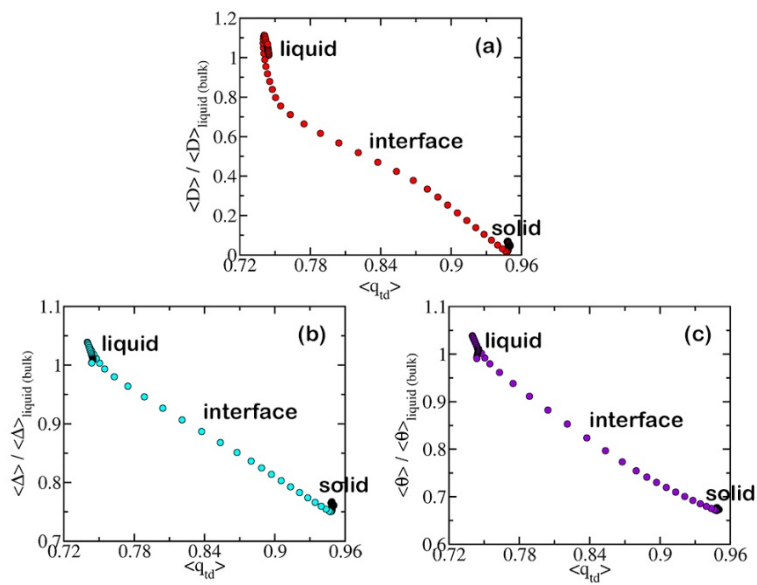

**Figure 6. Variation of dynamical variables in ice-water system as a function of the structural parameter $q_{td}$. (a) Diffusion coefficient (D), (b) Translational jump in 100 fs (Δ), and (c) Angular jump in 100 fs (θ). All dynamical variables are normalized by bulk values. Each point represents the average value of the concerned variables a grid along the z-direction. All the graphs show approximately proportional decay**.

The structural dependence of all the dynamical variables shows similar nature. The correlations between these parameters are determined by computing the Pearson correlation coefficient (**Eq. [14]**) for each of them against tetrahedrality order parameter.

$$\rho_{XY} = \frac{\text{cov}(X,Y)}{\sigma_X \sigma_Y} \quad [14]$$

where, 'cov' stands for covariance and $\sigma$ stands for standard deviation. $\rho_{XY}$ denotes the extent of correlation between any two variable X and Y. The values of correlation coefficient between all the dynamical variables and tetrahedrality order parameter are, ~-0.99 representing complete anti-correlation. The relation is almost linear. This signifies that for complete order, as in the ice-phase, dynamics is minimum. Transition to the liquid disordered state accelerates the system.

An interesting aspect of this correlation is the distribution of the points in the different regions of the two phase coexistent system. The distribution of $q_{td}$ in ice is a single sharp peak while it is much broader in liquid water. The dynamical features being essentially governed by the structure, shows similar characteristics. As a result, in **Figure 6** the points are much scattered in the liquid region than that in the solid. In the interfacial range, however, we do not find a significant number of points as it spans throughout a relatively small region with respect to the whole system.

In the present work, we have attempted to correlate the diffusion with the increasing tetrahedrality parameter as we travel from the liquid to the solid across the interface. The quasi-linear relation we observe numerically requires further investigation.

### B. Rectangular nano-confinement

As discussed earlier in the system and simulation details section, we keep the two hydrophobic graphene sheets fixed at 0 and 50 Å along z-axis, parallel to each other along the xy-plane. First, we investigate the structural modification of the confined water molecules along the inter planer axis (z) by evaluating the tetrahedrality order parameter ($q_{td}$) as shown in **Figure 7**. We observe that the convergence of $q_{td}$ to the bulk value is achieved within a short length scale of ~2-3 Å. The correlation length obtained from the spatial correlation function of $q_{td}$ (**Figure 7**, inset) is approximately 1.5 Å by using a single exponential function

**Eq. [9]** (n=1). Hence, under hydrophobic slab confinement, the perturbation in the tetrahedral structure due to surface effects is short ranged.

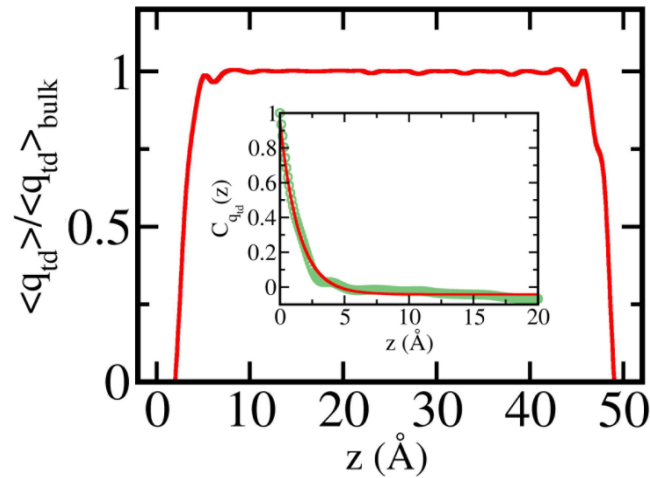

**Figure 7. Variation in the tetrahedrality order parameter ($q_{td}$) along z-axis (normalised to the bulk value). The two graphene sheets are placed at 0 and 50 Å. (inset) Spatial correlation of the tetrahedrality order parameter along z-axis, fitted with an exponential decay function.**

Next, we aim to obtain the dynamical length scales for this system. In view of that, we calculate the translational and rotational displacements within a fixed time interval of 100 fs as a measure of translational and rotational diffusion of the water molecules. We opt for such descriptions as the calculation of D from mean squared displacement or velocity autocorrelation is not possible because of statistically insignificant residence time in each of the bins. The translation and rotation profiles thus obtained, show oscillatory nature before their convergence toward the bulk value (**Figure 8a** and **8b**). We fit the spatial correlations of the translational and rotational motions along z by using linear combinations of damped exponentials (**Eq. [12]**; n=1). For translation we obtain two length scales of 0.9 Å (~96%) and 3.1 Å (~4%). In the case of rotation, the length scales obtained are 0.7 Å (~73%) and 2.6 Å (~27%) (**Figure 8c** and **8d**).

In order to probe the relation between structure and dynamics, we next plot translation and rotational mobility against tetrahedrality (**Figure 8e** and **8f**). It is clear from these two graphs that the increase in tetrahedrality reduces the mobility of the water molecules and becomes saturated after a critical value of $[<q_{td}>/<q_{td}>_{Bulk}] \sim 0.7$.

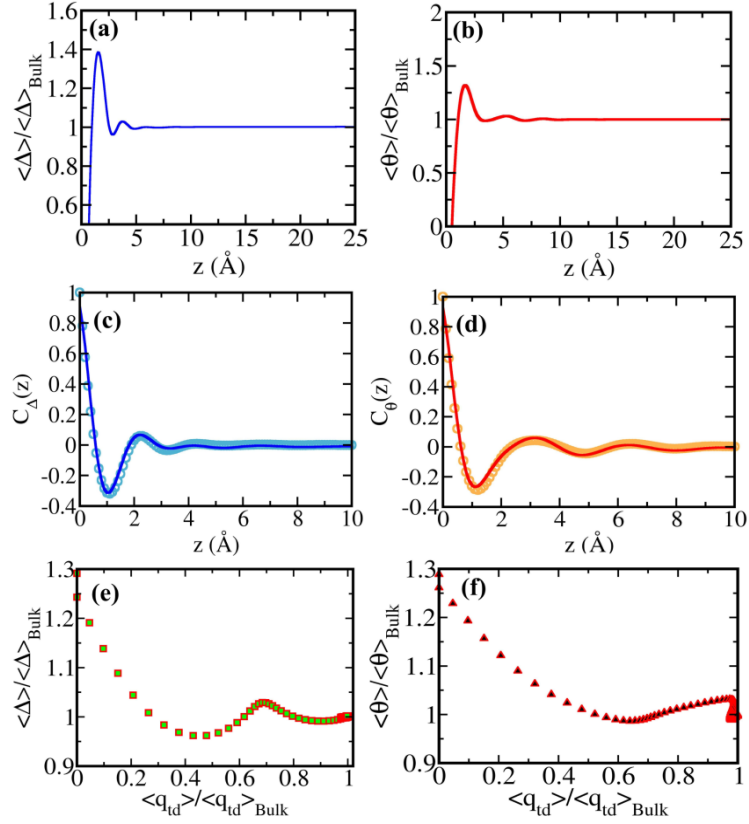

**Figure 8.** (a) Variation in the translational motion of water molecules along the inter slab axis (z-axis). The displacement of a water molecule within time intervals of 100 fs is considered as a measure of translational diffusion. (b) Variation in the rotational motion of water molecules along z-axis. In a similar fashion, the angular displacement within time intervals of 100 fs is considered as a measure of rotational diffusion. (c) and (d) are respectively the spatial correlations of translational displacement and rotational displacement profiles along z-axis. (e) and (f) are the correlation between structure and dynamics. That is, the variation of translational or rotational diffusion with tetrahedrality along the inter slab axis.

Another interesting collective quantity is the mean squared dipole moment fluctuations of the grids along the inter planer axis. This quantity is related to the static dielectric constant of the confined liquid.[83] Because of the geometry of the confinement, there arises an inherent anisotropy for the vectorial and tensorial properties. In this regard, we decompose the total dipole moment as a sum of parallel $\left(M_x + M_y\right)/2$ and perpendicular ($M_z$) components. In **Figure 9** we show the spatial variation of the parallel and perpendicular mean squared dipole moments. Interestingly, these two profiles behave quite differently. The parallel component exhibit a 3 fold increase near the surface whereas the perpendicular one shows a decrease by ~30%. This happens because of the propagation of opposite correlations from the surface and alignment of water molecules near the surface.

The correlation length obtained from the spatial correlation functions are as follows. We fit the oscillatory decay graphs (**Figure 9**, insets) by using a linear combination of a damped Gaussian and a damped exponential decay (**Eq. [13]**). For the parallel component, the length scales are 2.0 Å (~60%) and 3.5 Å (~40%); and for the perpendicular component the length scales obtained are 1.4 Å (~70%) and 2.1 Å (~30%).

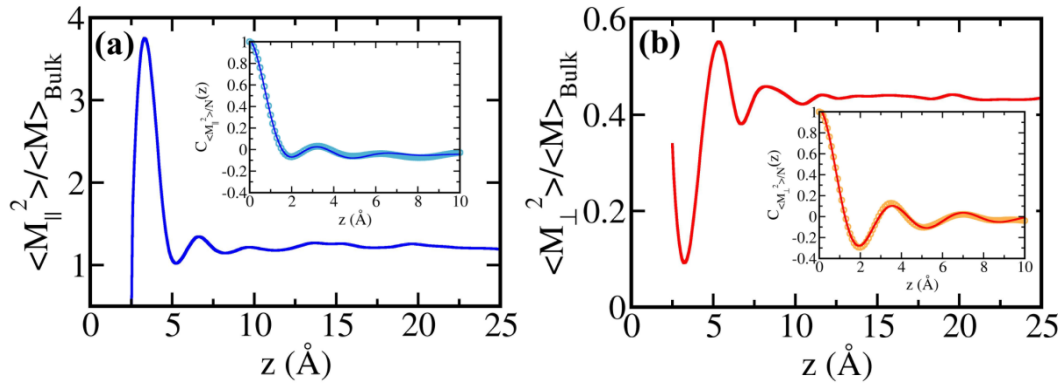

**Figure 9.** Spatial variation of (a) parallel ($M_x$ and $M_y$) and (b) perpendicular ($M_z$) components of the mean squared total dipole moment of different grids along z-axis. The values are normalised to the bulk value. Interestingly, near the surface, the parallel component becomes ~3 times higher, whereas the perpendicular component becomes ~30% lesser than that in the bulk. (insets) Spatial correlations of parallel and perpendicular dipole moments of the grids along the inter planer axis.

### C. Water on protein surface

Water on the protein surface is crucial for the proper functioning of the biomolecule. This system has been studied extensively, both by theory and experiments.[24, 84-86] By virtue of the heterogeneous nature of the protein surface, the interfacial water molecules often display a platter of relaxation timescales. Presence of the biomolecule disturbs the natural hydrogen bond network in water. This results in reorientation of these molecules which now form new hydrogen bonds with the protein residues. Hence, these water molecules, often referred to as *biological water*[10, 21] are structurally and dynamically different from bulk water.[25, 87] By studying the distributions of translational and orientational relaxation time scales, it was shown that protein hydration layers not only house slow water molecules, but also molecules faster than the bulk.[25] The timescales depend on the nature of the nearby protein residues.[30] Here, we investigate the structural and dynamical properties of water molecules in the vicinity of protein by fragmenting the water into 50 spherical bins of 1 Å width that radially spread out from the protein surface.

We choose the globular protein myoglobin for its approximately spherical shape. The structural arrangement of water molecules in each layer is determined by calculating the tetrahedrality order parameter $q_{td}$. We find that the water molecules near the protein surface, till about 7 Å shows lower average tetrahedrality. The tetrahedral arrangement being a consequence of the hydrogen bonding pattern in water, the presence of the protein surface severely affects the structure. $q_{td}$ shows an almost 12 % change as one goes radially outward from the protein surface. After about 10 Å, water attains bulk like structure (**Figure 10a**).

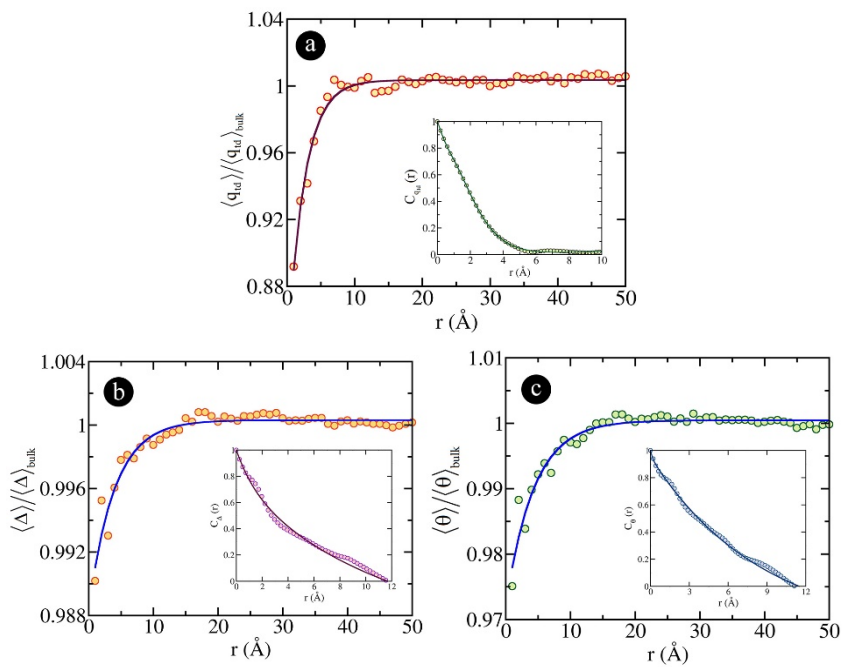

**Figure 10. Variation of structural and dynamical variables with distance from the protein surface. (a) tetrahedrality order parameter ($q_{td}$), (b) translational jump in 100 fs (D), and (c) angular jump in 100 fs (q). All the parameters are normalized with respect to bulk water values. The variation is similar in all the cases, with the dynamic variables attaining bulk like properties at a longer length scale. Insets show the spatial correlation of the respective variables. The relaxations are nonexponential displaying a multitude of length scales.**

The temperature of the system being high (300 K), the dynamics of water molecules is faster than the ice-water system (where T = 250 K). As a results the residence times of the water molecules in the 1 Å spherical shells is not high enough to calculate their mean square displacements for determination of diffusion coefficients. Hence, translational dynamics is estimated by the bin-averaged jumps executed by water molecules in a given time window (100 fs). Rotational dynamics is also characterized in the similar manner. (**Eqs. [3]** and **[4]**)

Dynamical variables also show related trend (**Figure 10b** and **8c**) like structure of water. Translational and rotational jumps (Δ and θ respectively) are smaller near the protein surface. However, they attain bulk-like values at lengths larger than that of $q_{td}$. To properly understand this difference in length scales between structural and dynamical behaviours of water on protein surface, we fit the data to exponential functions as defined in **Eq. [7]**. The fits are shown in **Figure 10**. The values of ξ obtained for $q_{td}$, Δ and θ are 2.4 Å, 3.9 Å and 4.3 Å respectively.

For a better insight into the relaxation of structure and dynamics for the interfacial water molecules, we calculate the spatial correlation function of the variables (**Eq. [6]**). We find that the relaxation of $q_{td}$ has power law dependence (**Eq. [8]**, n=2). The length scale associated with the power law is 1.01 Å with an exponent of 2.36. For the biexponential part, 50 % of the relaxation length is 0.5 Å while the other 50 % corresponds to 0.75 Å. The spatial correlation of both the dynamical variables can be efficiently described by biexponential functions (**Eq. [9]**, n=2). For translational jump, the length scales we obtain are 1.41 Å (18 %) and 11.9 Å (82 %). Similar parameters are also obtained for rotational jumps: 0.16 Å (4 %) and 8.58 Å (94 %). The fitted graphs are shown in the insets of the corresponding variables in **Figure 10**. The average relaxation length scales of Δ and θ are of the order of ~10 Å; whereas, that of $q_{td}$ is only ~0.5 Å. Hence, from both the aforementioned analyses, we see that protein surface perturbs the dynamics of water molecules to a greater extent than the corresponding structural organizations.

Collective response of water molecules can be evaluated from dipole moment averaged over water molecules in each shell. A particularly interesting feature is presented by the collective dipole moment time correlation function (TCF). The amplitude of the TCF $\langle M^2 \rangle$ gives an idea about the collective dynamical correlation of dipole moment in each shell. Hence spatial correlation of $\langle M^2 \rangle$ itself brings out the length scale of the effect of protein surface on the collective dynamics of water molecules.

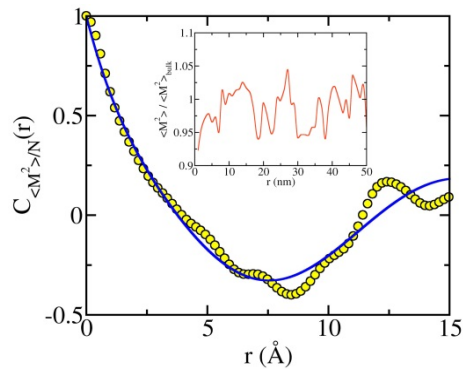

**Figure 11. Spatial correlation function of the value of shell averaged dipole moment TCF at time t = 0 ($\langle M^2 \rangle$) plotted against radial distance from the protein surface (r). Inset shows the value of $\langle M^2 \rangle$ as a function of r.**

The correlation function $C_{\langle M^2 \rangle/N}(t)$ shown in **Figure 11** is fitted to **Eq. [12]** for n = 2. The biexponential part of the relaxation provides the following two length scales: 1.41 Å (58 %) and 22.1 Å (42 %). In this collective dynamics as well, the relaxation length scale is of the order of ~10 Å. Hence ubiquitously, for water on protein surface we can say that dynamics is perturbed to a greater extent than structure. However, the nature of spatial relaxations of single molecule and collective properties are different.

### D. Aqueous DNA system

DNA is central to protein synthesis and gene expression. In order to do so, proteins (enzyme) need to bind to specific parts of the DNA. That is, molecular recognition becomes important. As the cytosol inside a cell is mostly water, it is important to know how far the water molecules might sense the presence of a protein/DNA. This, in turn, helps us to understand the water mediated interactions.[88] In this section we investigate the structural and dynamical correlations as a function of the distance (**r**) from the surface of the DNA.

To probe the structure of the water molecules, we calculate the tetrahedrality order parameter along **r** by considering bins of 1 Å width (**Figure 12**). However, the first few points should not be considered as a low tetrahedrality region as they are interfacial water molecules and not uniformly surrounded by other water molecule. We again observe a fast convergence of tetrahedrality with a correlation length of 1.6 Å (**Figure 12**, inset) by fitting the correlation with a single exponential function (**Eq. [9]**, n = 1).

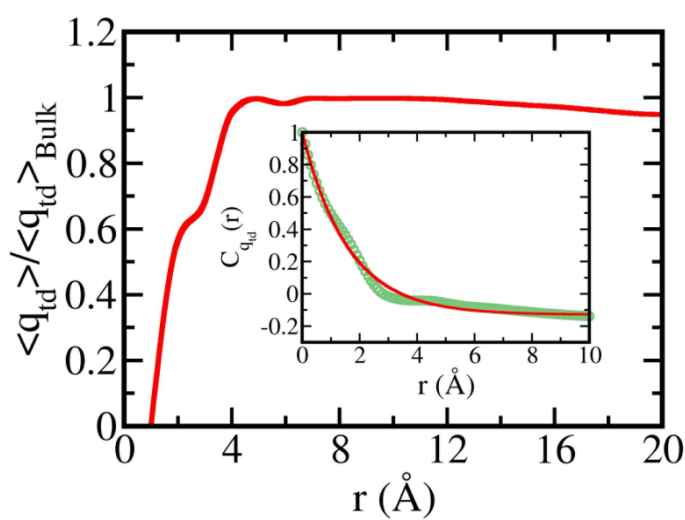

**Figure 12. Variation in the tetrahedrality order parameter ($q_{td}$) as a function of the distance from DNA surface (r), normalised to the bulk value. (inset) Spatial correlation of the tetrahedrality order parameter, fitted with an exponential decay function.**

On the other hand, the single particle dynamical correlations exhibit larger length scales. In **Figure 13a** and **5b,** we show the variation of translational and rotational mobility against distance **r**. The plots reveal an initial sluggishness in the dynamics followed by a short length scale oscillations and convergence. We observe an increase in the rotational mobility ~4-5 Å from the DNA surface. Similar observation has been reported in an earlier study where Bagchi *et al.* reported a broad log-normal distribution for rotational correlation timescales in the hydration layer of DNA.

The spatial correlations are fitted to a linear combination of Gaussian and exponential decay (**Eq. [13]**) (**Figure 13**, insets). The length scales of dynamical correlations obtained are as follows. For translational motion, the correlation lengths are 0.8 Å (~55%) and 12.1 Å (~45%). For rotational motion these are 1.23 Å (~73%) and 25.3 Å (~37%). These are surprisingly long ranged.

In **Figure 13c** and **13d** we show the relation between structure and dynamics by plotting the translational and rotational mobility against tetrahedrality. It shows a non-linear increase up to a critical value of $[<q_{td}>/<q_{td}>_{Bulk}]$ ~ 0.6, followed by an expected saturation. This indicates that an increase in the tetrahedrality increases mobility of the water molecules as a result of moving further away from the DNA surface. Both rotation and translation exhibits interesting cross-over behaviour as depicted in Figure 14(c) and 14(d). This seems to imply that the effects of DNA grooves sharply decrease at around 0.6 value of the bulk tetrahedrality parameter. This cross-over deserves further study.

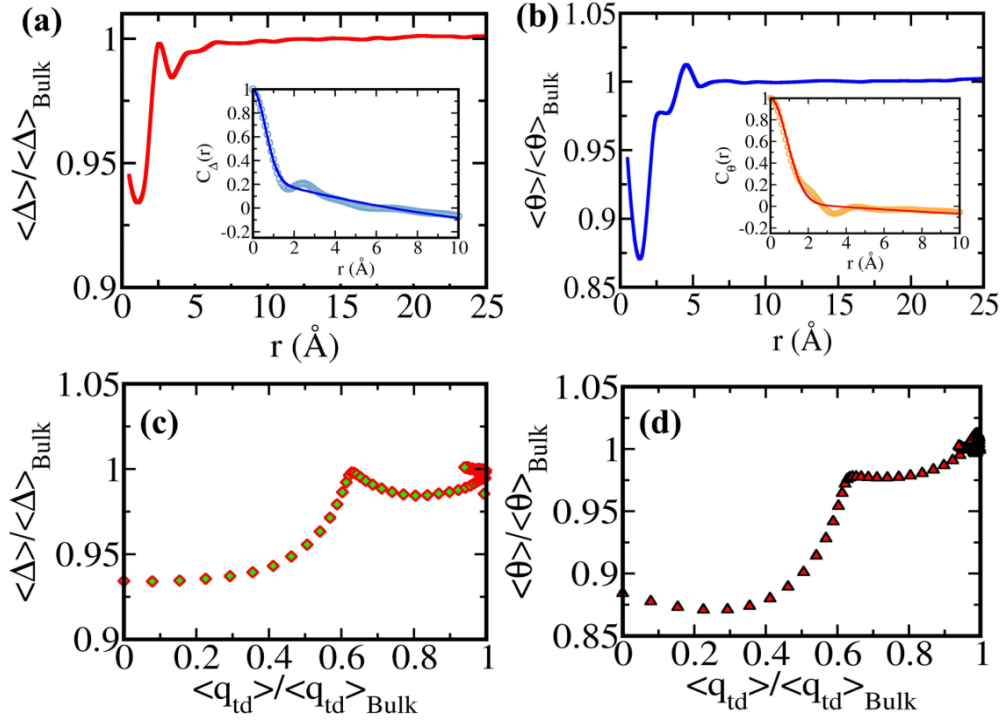

**Figure 13.** (a) Variation in the translational motion of water molecules as a function of the distance (r) from DNA surface. The displacement of a water molecule within time intervals of 100 fs is considered as a measure of translational diffusion. (b) Variation in the rotational motion of water molecules as a function of the distance (r) from DNA surface. The angular displacement of a water molecule within time intervals of 100 fs is considered as a measure of translational diffusion. (insets) The spatial correlation functions of the translational and rotational diffusion profiles shown in (a) and (b). (c) and (d) are the correlation between structure and dynamics. That is, the variation of translational or rotational diffusion with tetrahedrality.

Next, we aim to obtain the correlation length of mean square dipole moment fluctuations. We note that, up to 25Å the profile of $<M^2>$ retains non-bulk like value (**Figure 14a**). This indicates the long range effect on collective orientation. The spatial correlation function that corresponds to **Figure 14a** is fitted with a damped Gaussian decay Eq. [11] with $n=1$. The obtained correlation length scale is 4.4 Å.

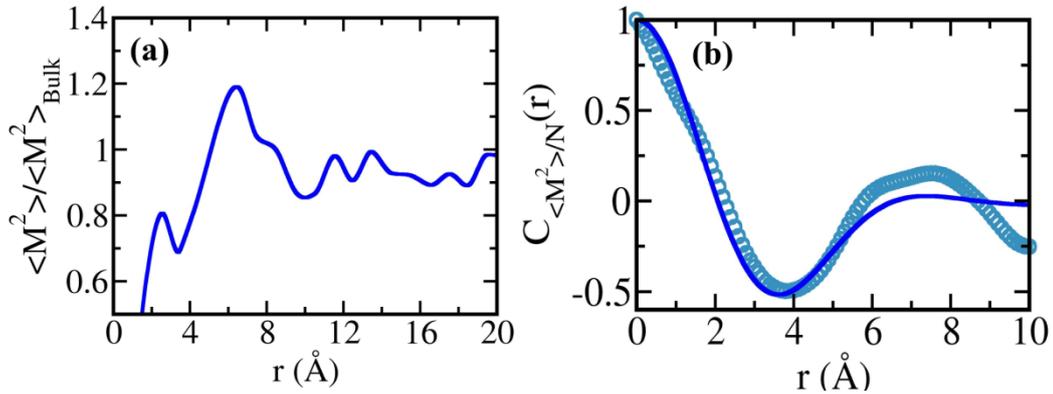

**Figure 14.** (a) Variation of the mean squared dipole moment as a function of the distance from the DNA surface. The oscillations do not converge for aqueous DNA system which indicates the presence of **long range perturbations. (b) Spatial correlation function of the mean squared dipole moment profile shown in (a).**

## V. CONCLUSION

A basic guiding principle in the study of liquid state properties is that "structure determines dynamics". For example, this has been the rationale behind the mode coupling theory approach to the liquid state properties and also glass transition.[89] This approach seems to work much of the time.

In the case of aqueous interfaces, the surface perturbs the structure of the surrounding water significantly. For an extended surface, the hydrogen bond network of water molecules gets terminated at the surface. Sometimes the water molecules form stronger hydrogen bonds with the surface groups (like in DNA and proteins), and other times there is no such enthalpic gain. In the latter case, as in the case of hydrophobic surface discussed here, water molecules attempt to minimize its free energy loss by preserving the maximum number of hydrogen bonds. This is turn requires certain orientation of the water molecules at the surface. This surface-induced orientation can propagate into the bulk, causing perturbation in the structure, and in turn, the dynamics which can be traced may be upto a few nm. The exact length scale is hard to find, and may depend on the surface and severity of perturbation.

One of the prime lessons of the mode coupling theory is that even a small change in structure can cause a large change in dynamics. This is well-known in the glass transition literature where the divergence of viscosity occurs due to a small increase in density or small lowering of temperature.[90-92] One can thus expect that small perturbations in water structure can cause large deviation in dynamical properties.

But water is a singular liquid. It is resistant to charge due to the presence of the extensive hydrogen bond network which renders it with a large effective force constant that again resists change. But it also can adopt many structural forms which are reflected in its large specific heat and large dielectric constant.

In accordance with our expectation, the dynamical perturbations propagate more into the bulk liquid than the static perturbations. This might also imply that one needs to look into more detailed characterization than the tetrahedrality parameter. We are presently studying these aspects.

We have earlier commented upon and presented analysis how single particle and collective dynamics differ at protein surface. This is certainly related to the elasticity of water which allows single particle dynamics to retain its bulk-like dynamical fastness. The collective properties like dielectric constant, and also density fluctuations as probed by scattering experiments, require longer distance into the bulk to recover the bulk-like characteristics.

We shall elsewhere present further analysis of the dynamical features of ice-water interface, and also water between hydrophobic surfaces. In particular, we shall explore the observed correlation between the diffusion constant and the tetrahedrality order parameter.

## Acknowledgements

We thank the Department of Science and Technology (DST, India) for partial support of this work. B. B. thanks Sir J. C. Bose fellowship for partial support. S. Mu. thanks DST, India for providing INSPIRE fellowship. S. Mo. thanks UGC, India for providing research fellowship.

## References

1. Bagchi, B. *Water in biological and chemical processes: From structure and dynamics to function*. Cambridge University Press, U.S.A.: 2013.
2. Ball, P. Life's matrix: water in the cell. *Cellular and molecular biology-Paris-Wegmann* **2001,** *47* (5), 717-720.
3. Laage, D.; Elsaesser, T.; Hynes, J. T. Water dynamics in the hydration shells of biomolecules. *Chem. Rev.* **2017,** *117* (16), 10694-10725.
4. Cho, M.; Fleming, G. R.; Saito, S.; Ohmine, I.; Stratt, R. M. Instantaneous normal mode analysis of liquid water. *J. Chem. Phys.* **1994,** *100* (9), 6672-6683.
5. Ohmine, I. Liquid water dynamics: collective motions, fluctuation, and relaxation. *J. Phys. Chem.* **1995,** *99* (18), 6767-6776.
6. Ohmine, I.; Saito, S. Water dynamics: Fluctuation, relaxation, and chemical reactions in hydrogen bond network rearrangement. *Acc. Chem. Res.* **1999,** *32* (9), 741-749.
7. Ohmine, I.; Tanaka, H. Fluctuation, relaxations, and hydration in liquid water. Hydrogen-bond rearrangement dynamics. *Chem. Rev.* **1993,** *93* (7), 2545-2566.
8. Sasai, M.; Ohmine, I.; Ramaswamy, R. Long time fluctuation of liquid water: 1/f spectrum of energy fluctuation in hydrogen bond network rearrangement dynamics. *J. Chem. Phys.* **1992,** *96* (4), 3045-3053.
9. Errington, J. R.; Debenedetti, P. G. Relationship between structural order and the anomalies of liquid water. *Nature* **2001,** *409* (6818), 318-321.
10. Pal, S. K.; Peon, J.; Bagchi, B.; Zewail, A. H. Biological water: femtosecond dynamics of macromolecular hydration. *J. Phys. Chem. B* **2002,** *106* (48), 12376-12395.
11. Mukherjee, S.; Mondal, S.; Deshmukh, A. A.; Gopal, B.; Bagchi, B. What Gives an Insulin Hexamer Its Unique Shape and Stability? Role of Ten Confined Water Molecules. *J. Phys. Chem. B* **2018,** *122* (5), 1631-1637.

12. Samanta, T.; Biswas, R.; Banerjee, S.; Bagchi, B. Study of distance dependence of hydrophobic force between two graphene-like walls and a signature of pressure induced structure formation in the confined water. *J. Chem. Phys.* **2018,** *149* (4), 044502.

13. Rasaiah, J. C.; Garde, S.; Hummer, G. Water in nonpolar confinement: from nanotubes to proteins and beyond. *Annu. Rev. Phys. Chem.* **2008,** *59*, 713-740.

14. Dokter, A. M.; Woutersen, S.; Bakker, H. J. Inhomogeneous dynamics in confined water nanodroplets. *Proc. Natl. Acad. Sci. U.S.A.* **2006,** *103* (42), 15355-15358.

15. Zheng, J.-m.; Chin, W.-C.; Khijniak, E.; Khijniak, E.; Pollack, G. H. Surfaces and interfacial water: Evidence that hydrophilic surfaces have long-range impact. *Adv. Colloid Interface Sci.* **2006,** *127* (1), 19-27.

16. Saito, S.; Ohmine, I. Dynamics and relaxation of an intermediate size water cluster ($H_2O$) 108. *J. Chem. Phys.* **1994,** *101* (7), 6063-6075.

17. Mondal, S.; Acharya, S.; Bagchi, B. Altered Dielectric Behaviour, Structure and Dynamics of Nanoconfined Dipolar Liquids: Signatures of Enhanced Cooperativity. *arXiv preprint arXiv:1904.05860* **2019**.

18. Pethig, R. Protein-water interactions determined by dielectric methods. *Annu. Rev. Phys. Chem.* **1992,** *43* (1), 177-205.

19. Grant, E. The dielectric method of investigating bound water in biological material: An appraisal of the technique. *Bioelectromagnetics* **1982,** *3* (1), 17-24.

20. Mashimo, S.; Kuwabara, S.; Yagihara, S.; Higasi, K. Dielectric relaxation time and structure of bound water in biological materials. *J. Phys. Chem.* **1987,** *91* (25), 6337-6338.

21. Nandi, N.; Bagchi, B. Dielectric relaxation of biological water. *J. Phys. Chem. B* **1997,** *101* (50), 10954-10961.

22. Nandi, N.; Bagchi, B. Anomalous dielectric relaxation of aqueous protein solutions. *J. Phys. Chem. A* **1998,** *102* (43), 8217-8221.

23. Nandi, N.; Bhattacharyya, K.; Bagchi, B. Dielectric relaxation and solvation dynamics of water in complex chemical and biological systems. *Chem. Rev.* **2000,** *100* (6), 2013-2046.

24. Bhattacharyya, S. M.; Wang, Z.-G.; Zewail, A. H. Dynamics of water near a protein surface. *J. Phys. Chem. B* **2003,** *107* (47), 13218-13228.

25. Mukherjee, S.; Mondal, S.; Bagchi, B. Distinguishing dynamical features of water inside protein hydration layer: Distribution reveals what is hidden behind the average. *J. Chem. Phys.* **2017,** *147* (2), 024901.

26. Mondal, S.; Mukherjee, S.; Bagchi, B. Decomposition of total solvation energy into core. side-chains and water contributions: Role of cross correlations and protein conformational fluctuations in dynamics of hydration layer. *Chem. Phys. Lett.* **2017,** *683*, 29-37.

27. Mukherjee, S.; Mondal, S.; Acharya, S.; Bagchi, B. DNA Solvation Dynamics. *J. Phys. Chem. B* **2018**.

28. Mondal, S.; Mukherjee, S.; Bagchi, B. Protein hydration dynamics: Much ado about nothing? *J. Phys. Chem. Lett.* **2017,** *8* (19), 4878-4882.

29. Mondal, S.; Mukherjee, S.; Bagchi, B. Protein Hydration Dynamics: Much Ado about Nothing? *J. Phys. Chem. Lett.* **2017,** *8* (19), 4878-4882.

30. Mondal, S.; Mukherjee, S.; Bagchi, B. Origin of diverse time scales in the protein hydration layer solvation dynamics: A simulation study. *J. Chem. Phys.* **2017,** *147* (15), 154901.

31. Ebbinghaus, S.; Kim, S. J.; Heyden, M.; Yu, X.; Heugen, U.; Gruebele, M.; Leitner, D. M.; Havenith, M. An extended dynamical hydration shell around proteins. *Proc. Natl. Acad. Sci. U.S.A.* **2007,** *104* (52), 20749-20752.

32. Johnson, M. E.; Malardier-Jugroot, C.; Murarka, R. K.; Head-Gordon, T. Hydration Water Dynamics Near Biological Interfaces. *J. Phys. Chem. B* **2009,** *113* (13), 4082-4092.

33.	Baldwin, R. L. Temperature dependence of the hydrophobic interaction in protein folding. *Proc. Natl. Acad. Sci. U.S.A.* **1986,** *83* (21), 8069-8072.
34.	Spolar, R. S.; Ha, J.-H.; Record, M. T. Hydrophobic effect in protein folding and other noncovalent processes involving proteins. *Proc. Natl. Acad. Sci. U.S.A.* **1989,** *86* (21), 8382-8385.
35.	Dill, K. A. Dominant forces in protein folding. *Biochemistry* **1990,** *29* (31), 7133-7155.
36.	Halle, B. Protein hydration dynamics in solution: a critical survey. *Philosophical Transactions of the Royal Society of London B: Biological Sciences* **2004,** *359* (1448), 1207-1224.
37.	Halle, B.; Nilsson, L. Does the dynamic Stokes shift report on slow protein hydration dynamics? *J. Phys. Chem. B* **2009,** *113* (24), 8210-8213.
38.	Mukherjee, S.; Mondal, S.; Acharya, S.; Bagchi, B. DNA Solvation Dynamics. *J. Phys. Chem. B* **2018,** *122* (49), 11743-11761.
39.	Huang, D. M.; Chandler, D. The hydrophobic effect and the influence of solute−solvent attractions. *J. Phys. Chem. B* **2002,** *106* (8), 2047-2053.
40.	Lum, K.; Chandler, D.; Weeks, J. D. Hydrophobicity at Small and Large Length Scales. *J. Phys. Chem. B* **1999,** *103* (22), 4570-4577.
41.	Hammer, M. U.; Anderson, T. H.; Chaimovich, A.; Shell, M. S.; Israelachvili, J. The search for the hydrophobic force law. *Faraday discussions* **2010,** *146*, 299-308.
42.	Israelachvili, J. N.; Adams, G. E. Measurement of forces between two mica surfaces in aqueous electrolyte solutions in the range 0–100 nm. *J. Chem. Soc., Faraday Trans. 1* **1978,** *74*, 975-1001.
43.	Meyer, E. E.; Rosenberg, K. J.; Israelachvili, J. Recent progress in understanding hydrophobic interactions. *Proc. Natl. Acad. Sci. U.S.A.* **2006,** *103* (43), 15739-15746.
44.	Despa, F.; Berry, R. S. The origin of long-range attraction between hydrophobes in water. *Biophys. J.* **2007,** *92* (2), 373-378.
45.	Choudhury, N. On the Manifestation of Hydrophobicity at the Nanoscale. *J. Phys. Chem. B* **2008,** *112* (20), 6296-6300.
46.	Samanta, T.; Bagchi, B. Temperature effects on the hydrophobic force between two graphene-like surfaces in liquid water. *J. Chem. Sci.* **2018,** *130* (3), 29.
47.	Li, J.; Morrone, J. A.; Berne, B. Are hydrodynamic interactions important in the kinetics of hydrophobic collapse? *J. Phys. Chem. B* **2012,** *116* (37), 11537-11544.
48.	Kurihara, K.; Kunitake, T. Submicron-range attraction between hydrophobic surfaces of monolayer-modified mica in water. *J. Am. Chem. Soc.* **1992,** *114* (27), 10927-10933.
49.	Senapati, S.; Chandra, A. Molecular relaxation in simple dipolar liquids confined between two solid surfaces. *Chem. Phys.* **1998,** *231* (1), 65-80.
50.	Senapati, S.; Chandra, A. Computer simulations of dipolar liquids near charged solid surfaces: electric-field-induced modifications of structure and dynamics of interfacial solvent. *J. Mol. Struct.: THEOCHEM* **1998,** *455* (1), 1-8.
51.	Louden, P. B.; Gezelter, J. D. Friction at Ice-Ih/Water Interfaces Is Governed by Solid/Liquid Hydrogen-Bonding. *J. Phys. Chem. C* **2017,** *121* (48), 26764-26776.
52.	Sanz, E.; Vega, C.; Abascal, J.; MacDowell, L. Phase diagram of water from computer simulation. *Phys. Rev. Lett.* **2004,** *92* (25), 255701.
53.	Salzmann, C. G. Advances in the experimental exploration of water's phase diagram. *J. Chem. Phys.* **2019,** *150* (6), 060901.
54.	Hayward, J. A.; Haymet, A. The ice/water interface: Molecular dynamics simulations of the basal, prism,{2021}, and {2110} interfaces of ice Ih. *J. Chem. Phys.* **2001,** *114* (8), 3713-3726.

55. Nada, H.; Furukawa, Y. Anisotropy in growth kinetics at interfaces between proton-disordered hexagonal ice and water: A molecular dynamics study using the six-site model of H2O. *J. Cryst. Growth* **2005,** *283* (1-2), 242-256.
56. Karim, O. A.; Haymet, A. The ice/water interface: A molecular dynamics simulation study. *J. Chem. Phys.* **1988,** *89* (11), 6889-6896.
57. Karim, O. A.; Haymet, A. The ice/water interface. *Chem. Phys. Lett.* **1987,** *138* (6), 531-534.
58. Bagchi, B. Self-diffusion across the liquid–crystal interface. *J. Chem. Phys.* **1985,** *82* (12), 5677-5684.
59. Matsumoto, M.; Yagasaki, T.; Tanaka, H. GenIce: Hydrogen-Disordered Ice Generator. *J. Comput. Chem.* **2018,** *39* (1), 61-64.
60. Abascal, J.; Sanz, E.; García Fernández, R.; Vega, C. A potential model for the study of ices and amorphous water: TIP4P/Ice. *J. Chem. Phys.* **2005,** *122* (23), 234511.
61. Berendsen, H.; Grigera, J.; Straatsma, T. The missing term in effective pair potentials. *J. Phys. Chem.* **1987,** *91* (24), 6269-6271.
62. Senapati, S.; Chandra, A. Dielectric constant of water confined in a nanocavity. *J. Phys. Chem. B* **2001,** *105* (22), 5106-5109.
63. Mukherjee, S.; Mondal, S.; Bagchi, B. Mechanism of Solvent Control of Protein Dynamics. *Phys. Rev. Lett.* **2019,** *122* (5), 058101.
64. Bernstein, F. C.; Koetzle, T. F.; Williams, G. J.; Meyer, E. F.; Brice, M. D.; Rodgers, J. R.; Kennard, O.; Shimanouchi, T.; Tasumi, M. The protein data bank. *European Journal of Biochemistry* **1977,** *80* (2), 319-324.
65. Oostenbrink, C.; Villa, A.; Mark, A. E.; Van Gunsteren, W. F. A biomolecular force field based on the free enthalpy of hydration and solvation: the GROMOS force-field parameter sets 53A5 and 53A6. *J. Comput. Chem.* **2004,** *25* (13), 1656-1676.
66. Case, D. A.; Cheatham III, T. E.; Darden, T.; Gohlke, H.; Luo, R.; Merz Jr, K. M.; Onufriev, A.; Simmerling, C.; Wang, B.; Woods, R. J. The Amber biomolecular simulation programs. *J. Comput. Chem.* **2005,** *26* (16), 1668-1688.
67. Lindorff-Larsen, K.; Piana, S.; Palmo, K.; Maragakis, P.; Klepeis, J. L.; Dror, R. O.; Shaw, D. E. Improved side-chain torsion potentials for the Amber ff99SB protein force field. *Proteins:Struct. Funct. Bioinfo.* **2010,** *78* (8), 1950-1958.
68. Jorgensen, W. L.; Chandrasekhar, J.; Madura, J. D.; Impey, R. W.; Klein, M. L. Comparison of simple potential functions for simulating liquid water. *J. Chem. Phys.* **1983,** *79* (2), 926-935.
69. Abraham, M. J.; Murtola, T.; Schulz, R.; Páll, S.; Smith, J. C.; Hess, B.; Lindahl, E. GROMACS: High performance molecular simulations through multi-level parallelism from laptops to supercomputers. *SoftwareX* **2015,** *1*, 19-25.
70. Press, W. H.; Teukolsky, S. A.; Vetterling, W. T.; Flannery, B. P. *Numerical Recipes 3rd Edition: The Art of Scientific Computing*. Cambridge University Press: 2007; p 1256.
71. Allen, M. P.; Tildesley, D. J. *Computer simulation of liquids*. Oxford university press: 2017.
72. Frenkel, D.; Smit, B. *Understanding molecular simulation: from algorithms to applications*. Elsevier: 2001; Vol. 1.
73. Nosé, S. A unified formulation of the constant temperature molecular dynamics methods. *J. Chem. Phys.* **1984,** *81* (1), 511-519.
74. Hoover, W. G. Canonical dynamics: equilibrium phase-space distributions. *Physical review A* **1985,** *31* (3), 1695.
75. Parrinello, M.; Rahman, A. Polymorphic transitions in single crystals: A new molecular dynamics method. *Journal of Applied physics* **1981,** *52* (12), 7182-7190.

76. Hess, B.; Bekker, H.; Berendsen, H. J.; Fraaije, J. G. LINCS: a linear constraint solver for molecular simulations. *J. Comput. Chem.* **1997,** *18* (12), 1463-1472.
77. Darden, T.; York, D.; Pedersen, L. Particle mesh Ewald: An N· log (N) method for Ewald sums in large systems. *J. Chem. Phys.* **1993,** *98* (12), 10089-10092.
78. Petrenko, V. F.; Whitworth, R. W. *Physics of ice*. OUP Oxford: 1999.
79. Errington, J. R.; Debenedetti, P. G.; Torquato, S. Cooperative Origin of Low-Density Domains in Liquid Water. *Phys. Rev. Lett.* **2002,** *89* (21), 215503.
80. Matsumoto, M.; Saito, S.; Ohmine, I. Molecular dynamics simulation of the ice nucleation and growth process leading to water freezing. *Nature* **2002,** *416* (6879), 409.
81. Mochizuki, K.; Matsumoto, M.; Ohmine, I. Defect pair separation as the controlling step in homogeneous ice melting. *Nature* **2013,** *498*, 350.
82. Bagchi, B. *Statistical Mechanics for Chemistry and Materials Science*. CRC Press, Taylor & Francis Group, New York, U.S.A.: 2018.
83. Gekle, S.; Netz, R. R. Anisotropy in the dielectric spectrum of hydration water and its relation to water dynamics. *J. Chem. Phys.* **2012,** *137* (10), 104704.
84. Bhattacharyya, K. Nature of biological water: a femtosecond study. *Chem. Commun.* **2008,** (25), 2848-2857.
85. Bagchi, B. Water dynamics in the hydration layer around proteins and micelles. *Chem. Rev.* **2005,** *105* (9), 3197-3219.
86. Bagchi, K.; Roy, S. Sensitivity of Water Dynamics to Biologically Significant Surfaces of Monomeric Insulin: Role of Topology and Electrostatic Interactions. *J. Phys. Chem. B* **2014,** *118* (14), 3805-3813.
87. Ghosh, R.; Banerjee, S.; Hazra, M.; Roy, S.; Bagchi, B. Sensitivity of polarization fluctuations to the nature of protein-water interactions: Study of biological water in four different protein-water systems. *J. Chem. Phys.* **2014,** *141* (22), 22D531.
88. Janin, J. Wet and dry interfaces: the role of solvent in protein–protein and protein–DNA recognition. *Structure* **1999,** *7* (12), R277-R279.
89. Bagchi, B.; Bhattacharyya, S. Mode coupling theory approach to liquid state dynamics. *Adv. Chem. Phys.* **2001,** *116*, 67-222.
90. Angell, C. A. Perspective on the glass transition. *J. Phys. Chem. Solids* **1988,** *49* (8), 863-871.
91. Mauro, J. C.; Yue, Y.; Ellison, A. J.; Gupta, P. K.; Allan, D. C. Viscosity of glass-forming liquids. *Proc. Natl. Acad. Sci. U.S.A* **2009,** *106* (47), 19780-19784.
92. Das, S. P.; Mazenko, G. F.; Ramaswamy, S.; Toner, J. J. Hydrodynamic Theory of the Glass Transition. *Phys. Rev. Lett.* **1985,** *54* (2), 118-121.